\documentclass{article}
\usepackage{icmc2025_paper_template}
\usepackage{times}
\usepackage{ifpdf}
\usepackage{soul}
\usepackage[english]{babel}
\usepackage{amsmath}

\def\papertitle{Expressive Music Data Processing and Generation}
\def\firstauthor{Jingwei Liu}
\def\secondauthor{Second Author}
\def\thirdauthor{Third Author}

\newif\ifpdf
\ifx\pdfoutput\relax
\else
   \ifcase\pdfoutput
      \pdffalse
   \else
      \pdftrue
  \fi
\fi

\ifpdf 
  \usepackage[pdftex,
    pdftitle={\papertitle},
    pdfauthor={\firstauthor, \secondauthor, \thirdauthor},
    bookmarksnumbered, 
    pdfstartview=XYZ 
   ]{hyperref}

  \usepackage[pdftex]{graphicx}
  \graphicspath{{./figures/}}
  \DeclareGraphicsExtensions{.pdf,.jpeg,.png}

  \usepackage[figure,table]{hypcap}

\else 
  \usepackage[dvips,
    bookmarksnumbered, 
    pdfstartview=XYZ 
  ]{hyperref}  

  \usepackage[dvips]{epsfig,graphicx}
  \graphicspath{{./figures/}}
  \DeclareGraphicsExtensions{.eps}

  \usepackage[figure,table]{hypcap}
\fi

\hypersetup{
    colorlinks,%
    citecolor=black,%
    filecolor=black,%
    linkcolor=black,%
    urlcolor=black
}

\title{\papertitle}

\oneauthor
  {\firstauthor} {University of California San Diego \\ %
    {\tt \href{mailto:jil121@ucsd.edu}{jil121@ucsd.edu}}}

\begin{document}
\capstartfalse
\maketitle
\capstarttrue
\begin{abstract}

Musical expressivity and coherence are indispensable in music composition and performance, while often neglected in modern AI generative models. In this work, we introduce a listening-based data-processing technique that captures the expressivity in musical performance. This technique derived from Weber's law reflects the human perceptual truth of listening and preserves musical subtlety and expressivity in the training input. To facilitate musical coherence, we model the output interdependencies among multiple arguments in the music data such as pitch, duration, velocity, etc. in the neural networks based on the probabilistic chain rule. In practice, we decompose the multi-output sequential model into single-output submodels and condition previously sampled outputs on the subsequent submodels to induce conditional distributions. Finally, to select eligible sequences from all generations, a tentative measure based on the output entropy was proposed. The entropy sequence is set as a criterion to select predictable and stable generations, which is further studied under the context of informational aesthetic measures to quantify musical pleasure and information gain along the music tendency\footnote{Please access the code and generated samples at the GitHub page: \href{https://github.com/Jovie-Liu/Expressive-Piano-MIDI-Generation}{https://github.com/Jovie-Liu/Expressive-Piano-MIDI-Generation}}.

\end{abstract}

\section{Introduction}\label{sec:introduction}

Computer-based music generation starts from computer-assisted composition\cite{assayag1999computer}, algorithmic composition\cite{nierhaus2009algorithmic}, until statistical machine learning and AI generative systems\cite{dubnov2023deep}. Deep learning and AI methods are taking over music generation and becoming the main tools for exploration nowadays \cite{briot2017deep, fernandez2013ai, ji2020comprehensive}. Among all the techniques and problems, we find that \textit{musical expressivity} is often ignored or less attended to both when preparing music data for training and designing generative architectures and sampling methods for neural networks \cite{cancino2018computational}. Besides, as the music generation is treated more like an engineering problem nowadays, we consider a lack of musical knowledge and insights into the generative process will cause a loss of musical intuition thus impeding the model from realizing better musical performance.

In this paper, we will first introduce an innovative listening-based data-processing technique that captures the expressivity in music performance in Section \ref{sec2}. In Section \ref{sec3}, we will deliberate on the multi-argument nature of musical systems and study how to model the interdependencies among output arguments under neural network structures. By designing a generative model capturing output interdependency, we better encode musical logic and intuition into neural networks, producing musically coherent outputs. In Section \ref{sec4}, we propose an innovative screening criterion to select eligible sequences from all generations. The output entropy is used to measure the predictability and stability of generated samples, and the statistics of the entropy sequence are investigated under the context of musical information dynamics and informational aesthetic measures to reveal inherent musical properties from the generated sequences of the neural network model.

\section{Listening-based Data Processing}\label{sec2}

The dataset we use in this model is the Yamaha e-Piano Competition MIDI dataset \cite{performance-rnn-2017}, a single-instrument MIDI format dataset recorded live piano performances. MIDI is a digital symbolic music representation format. It serves as an efficient mid-level representation of audio with reduced size, which also provides more freedom than musical staff when making notes of music. This dataset contains over 1000 MIDI captures of real-time piano performances with high precision in time and dynamics by skilled pianists. The performance expressivity is well preserved in the MIDI data, which can be reproduced when we auralize these MIDI files.

One objective of processing these data as inputs to the neural networks is to minimize the information loss and preserve its expressivity. As the MIDI format doesn't impose fixed grids and quantized durations in metrics, we are flexible to work with time in milliseconds, thus the expressive micro-timing is encoded in the musical representations. Due to the fine scale of time, one crucial technical difficulty in symbolic music generation -- chords and polyphony, which demand the generation of vertically disposed notes, are naturally resolved as we notice that there is no actual simultaneous notes. No matter how close the onsets of two notes are, there is always a time discrepancy between them. In other words, the onset of notes can always be scheduled sequentially as an ordered topology. In this sense, autoregressive neural networks are eligible to generate symbolic music with rich polyphonic texture under a fine time scale.

\begin{figure}[h]
\centering
\includegraphics[width=1\columnwidth]{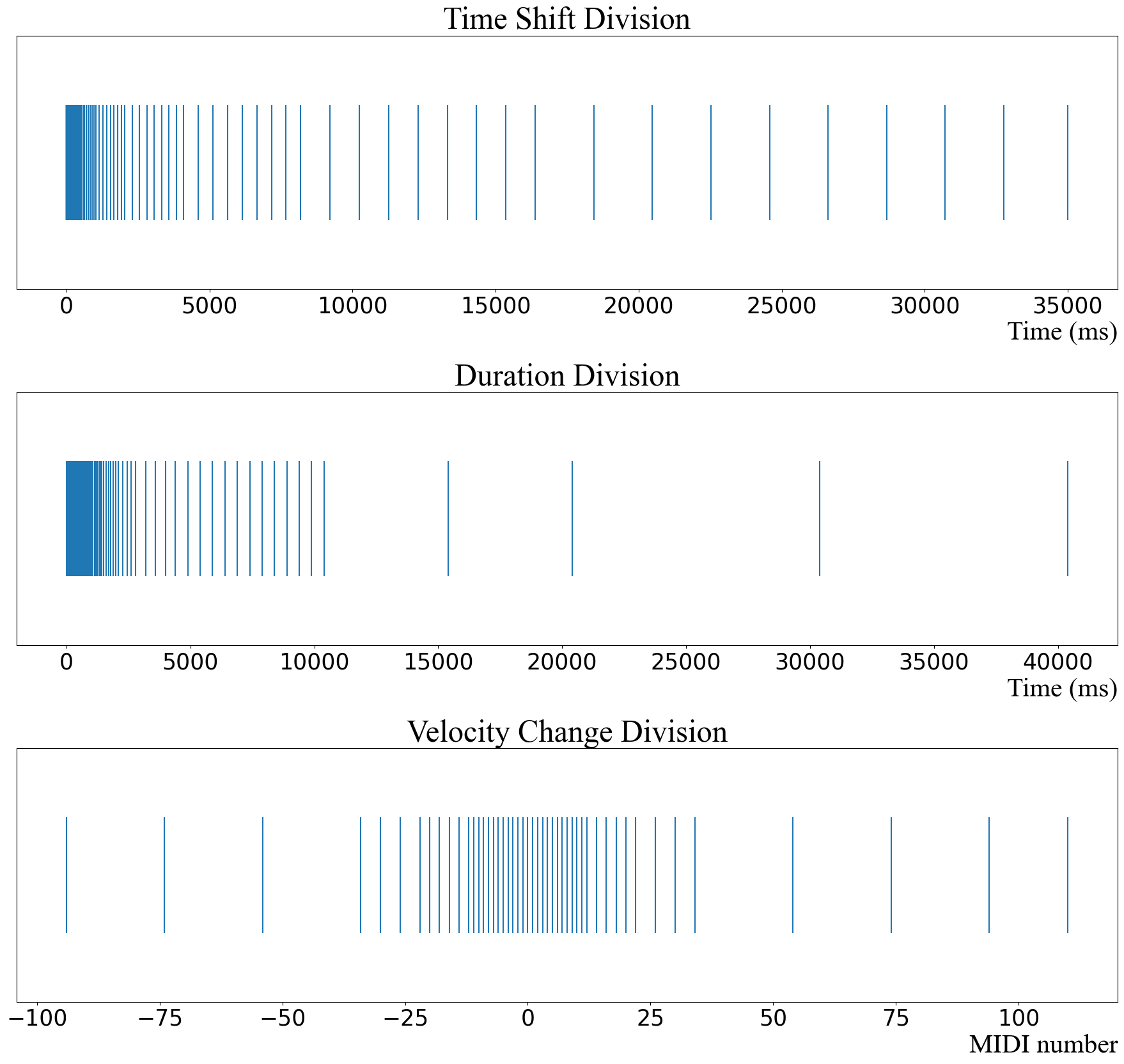}
\caption{Categorical Distributions of Time Shift, Duration, and Velocity Change. The divisions are co-determined with Weber's law where the perceptual changes are proportional to current values, and the ground truth statistics to balance the data distributions in each of the training classes.
\label{fig: division}}
\end{figure}

The expressive timing and dynamics are processed with quasi-mel-scale divisions as shown in Figure \ref{fig: division}. These intervals are determined in two folds. Taking account of the human perceptual truth depicted by Weber's law, the just noticeable difference is proportional to the current value. For example, for two subsequent notes with a time shift of $10ms$, increasing it to $15ms$ should be easily perceived; however, if the time shift is $1s$, adding $5ms$ makes little difference perceptually, thus a categorical interval of $[1s, 1.005s]$ doesn't make sense, the intervals should be expanded as the values increase.

The same rules apply to the duration, which is also measured in milliseconds. As for the dynamics, MIDI categorizes the velocity of pressing the keys into $128$ discrete steps, with loudness increasing from $0$ to $127$ (the quantization here is already logarithmic w.r.t. loudness perception, e.g. decibels). Here we also apply the perceptual change rules in processing the velocity values. The assumption is that, instead of absolute value, the listeners are more sensitive to the changes in dynamics. Therefore, we replace the velocity term with velocity change that encodes the dynamic difference between subsequent notes. Again, we apply Weber's law to the categorization of velocity change intervals, where the division steps are proportional to the absolute velocity change values (Figure \ref{fig: division}).

As mentioned above, the categorical intervals follow a quasi-mel scale. Weber's law applies at large but the intervals do not necessarily expand exponentially. To determine the fine scales of the divisions, on the other hand, we perform a statistical analysis of the dataset to extract the distributions of time shift, duration, and velocity change in the ground truth. When dividing the training classes, we balance the frequencies of data inputs in each class so that the model can be trained more efficiently with evenly distributed output classes.

As for note pitches, we follow the default MIDI note of $128$ values. In our dataset, piano notes range from $21$ to $108$. Besides note events, we also take care of control events that have profound impacts on musical perception such as the sustain pedal in piano performance. Instead of $128$ MIDI pedal values, we notate this event simply with on/off (1/0) to represent the pedal status for each note (e.g. value $\ge 64$ is on, otherwise is off). The terms and results of data processing are summarized in Table \ref{tab1}.

\begin{table}[h]
 \begin{center}
 \begin{tabular}{|l|l|l|}
  \hline
  Arguments & MIDI Info & Neural Network I/O\\
  \hline
  Note value ($n$)& $[21,108]$ & $88$ classes $[0,87]$\\
  \hline
   Time shift ($t$)& time in $ms$ & $105$ classes $[0,104]$\\
  \hline
   Duration ($d$)& time in $ms$ & $120$ classes $[0,119]$\\
  \hline
   Velocity ($v$)& $[0,127]$ & velocity change $[0,46]$\\
  \hline
   Pedal status ($p$)& $[0,127]$ & binary states $\{0,1\}$\\
  \hline
 \end{tabular}
\end{center}
 \caption{MIDI Data Processing as Neural Network Input/Output}
 \label{tab1}
\end{table}

\section{Multi-arguments Sequential Model}\label{sec3}
As discussed in Section \ref{sec2}, there are five input/output arguments for each note. Before devising a generative model, we need to understand the intercorrelations among these arguments. 
\subsection{Related Work}
This work \cite{yang2019deep} performed a disentanglement of pitch and rhythm and argued that the pitch contour or rhythmic patterns of one piece can be imposed on another piece thus generating new music. This assumption implicitly treats pitch and rhythm as two independent components in a piece, which can be refuted with many musical examples. The relationships among musical elements are complicated. On the one hand, any separate study of a single element in a piece, such as pitch, rhythm, dynamics, tempo, etc. without considering the effects of all other elements is unjustified. Music is an inseparable entity and everything is finely coordinated within the context. On the other hand, these elements seem to have autonomy to certain degrees. For example, in practice, some rhythmic patterns are reused for various melodic lines, and the same pitch contour can be performed at different rhythms and tempos for contrasting effects. However, such operations require complicated musical knowledge that's not yet programmed or trained in the machines.

Another example is the multitrack music transformer \cite{dong2023multitrack}, where the multiple output fields are modeled independently. Although the author pointed out the limitations and advantages of their model, musically speaking, it's logically inconsistent in practice. As the generative model is probabilistic, modeling output arguments independently means the choice in one argument has no impact on the decisions in other arguments. In general musical cases, if a tone is dissonant, it's usually presented shorter so the music can be passed to the next structural note smoothly. Also, when a tone is presented with a short duration, it's usually performed lighter with little emphasis. The independent model violates lots of musical intuitions and fails to capture the intercorrelation among various musical elements. The upper model in Figure \ref{fig: LSTM5} reflects this structure, with falsely attributed output distribution:
\begin{equation}
    p(x_n, x_t,x_d,x_v,x_p) \to  p(x_n) p(x_t) p(x_d) p(x_v) p(x_p) \nonumber
\end{equation}
where the subscripts denote the argument types specified in Table \ref{tab1}.

\subsection{Model with Interdependency}

The distribution of output can be decomposed with chain rules:
\begin{align}
    p(x_n, x_t,x_d,x_v,& x_p)  = p(x_n) \cdot p(x_t|x_n) \cdot p(x_d|x_n,x_t) \nonumber \\
    & \cdot p(x_v|x_n,x_t,x_d) \cdot p(x_p|x_n,x_t,x_d,x_v) \nonumber
\end{align}
It shows us the plausibility of sampling the output arguments one after another, with distributions conditioned on all previously sampled arguments of this note. This type of conditioning captures full interdependencies among all arguments. However, modeling such interdependency using neural networks demands a redesign of the model structure.

In generating musical sequences, we naturally chose autoregressive neural networks. We didn't use the transformer here because it's computationally expensive (the model is trained on the author's personal computer with one GPU). That being said, modeling multi-argument interdependency using transformers also raises unique challenges, which is out of the scope of this paper.


In autoregressive neural networks, the next note is generated with the current input and the immediate previous hidden state propagated along the sequence. When there are multiple output arguments and the current argument distribution is conditioned on the previous argument samples, the output device tends to be involved. One possible solution is to stretch the multi-argument outputs into a sequence and convolve the hidden states with these arguments autoregressively, with the previously sampled argument as the next input. This treatment has lots of drawbacks and the most prominent one is the extended sequence length. Autoregressive neural networks suffer from vanishing gradients and have difficulty dealing with long sequences due to limited memory, so extending the note sequence five times longer will degrade the model performance. Another option is to parallel the output arguments, namely using the same hidden states to compute all arguments. However, unlike in the independent case, here we add one more step to take the previously sampled arguments as input to generate the current argument distribution with the hidden states. This treatment also has its problems. First, we assume the hidden states encode memories for all five arguments. As discussed before, the information for generating pitch and rhythm might be disentangled at certain levels. In other words, the system may refer to totally different features when making decisions for the next note's pitch and duration, and the processes of integrating information for predicting these two arguments might also be widely apart. Cramming all features of various musical elements into a single hidden state may increase prediction errors. Secondly, the influence of other arguments on the current one is context-dependent. For example, when a rhythmic pattern is imperative at force, the duration argument can choose to ignore the conditioned pitch value. However, in another case when the pitch argument sampled a dissonant singular note, the duration argument might respond strongly to sample a short duration. In other words, the conditioning effects of other output arguments should be soft-coded and assigned respective weights according to the context.

We designed such a neural network that partially captures the autonomy of musical elements (disentangled hidden states for different arguments) and processes conditioning with a local attentional device\cite{liu2024expressive}. This neural network follows a quasi-LSTM structure with internal disentangled attention mechanisms, thus named the LSTM-Attention model. The author believes the devising idea of this neural network tackles lots of problems mentioned above. However, as the internal connections of the cell are too complicated and many extra parameters are involved, the backpropagation through time is stagnated thus leaving the neural network barely trained. This is another drawback of parallel multi-argument computation with conditioning. The shared hidden states are involved in multiple subprocesses and extra parameters are added to condition the input, which makes the backpropagation of derivatives unbalanced thus impeding training and convergence.

\begin{figure}[h]
\centering
\includegraphics[width=0.9\columnwidth]{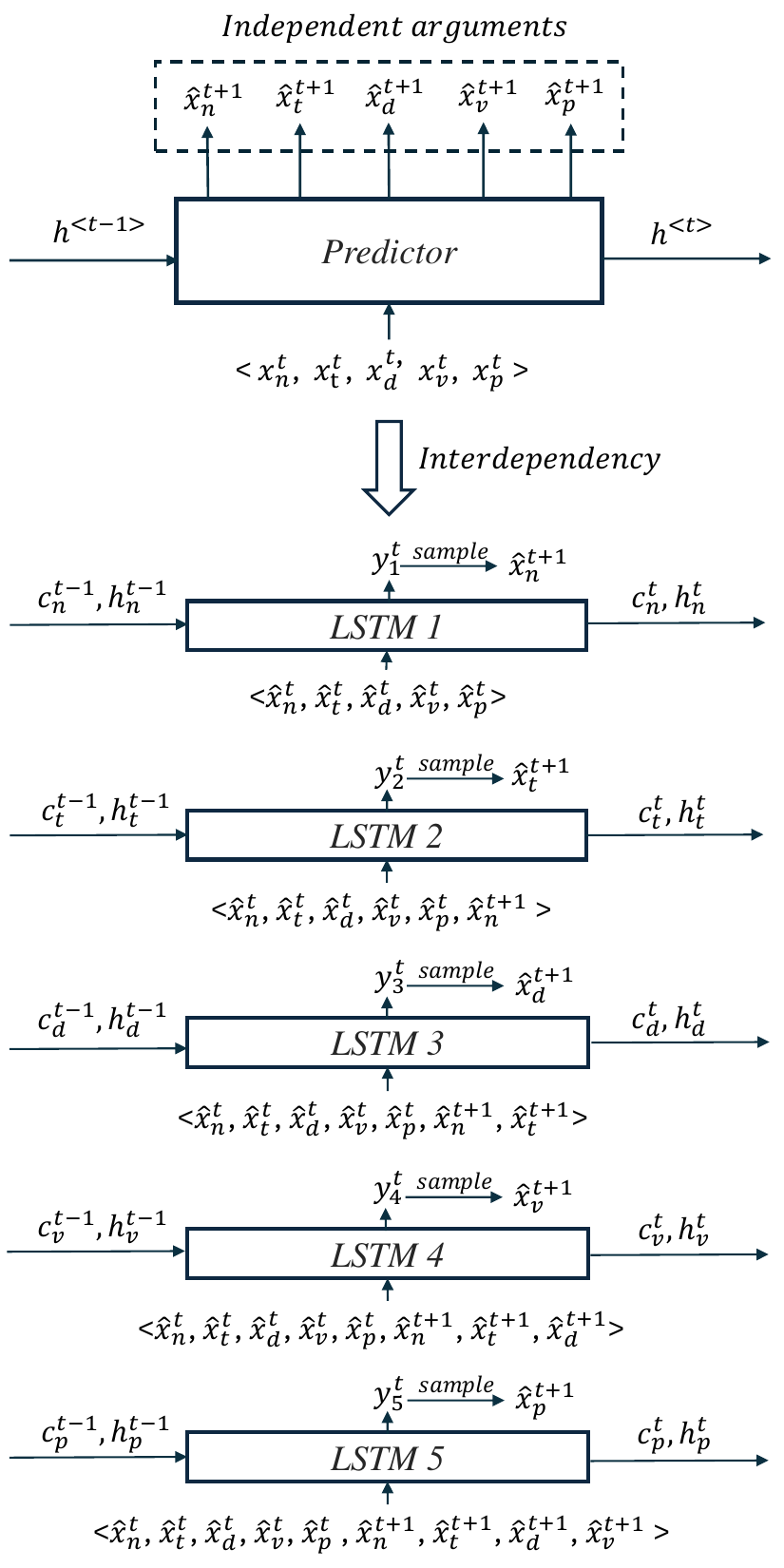}
\caption{A Way to Capture Interdependency in a Multi-argument Sequential Model. The interdependencies are modeled with probabilistic conditioning and the multi-argument output model is decomposed into separate sequential submodels with a single output.} \label{fig: LSTM5}
\end{figure}

It seems difficult to design a perfect structure to handle all the problems of multi-argument interdependencies in a single autoregressive neural network. However, if we take a step back and stop being so insistent on packing all functions in one neural network, there is a plausible option with computational trade-offs. As shown in Figure \ref{fig: LSTM5}, we can decompose the multi-argument output neural network into multiple neural networks with a single output. Here the hidden states are automatically disentangled for each argument to encode information relevant to the prediction of that argument specifically. The sampled arguments are conditioned in the next neural network as appended input, which avoids the excessive treatment of conditioning and sophisticated neural network structures.

\subsection{Attention Score}

The sequential model we use is the Long Short-Term Memory (LSTM) neural network, a recurrent neural network with forgetting mechanisms that can selectively update cell memory based on the current input. The LSTMs are built in PyTorch with $2$ stacked layers and of hidden size $150$. When the complexity of an integrated model is disseminated into several submodels, the trade-off kicks in as more computations are needed to compensate for the reduced complexity. However, the computations do not increase fivefold since with simpler submodels we can effectively reduce the latent dimensions for faster convergence. The five LSTMs are trained separately as independent neural networks.

\begin{figure}[h]
\centering
\includegraphics[width=\columnwidth]{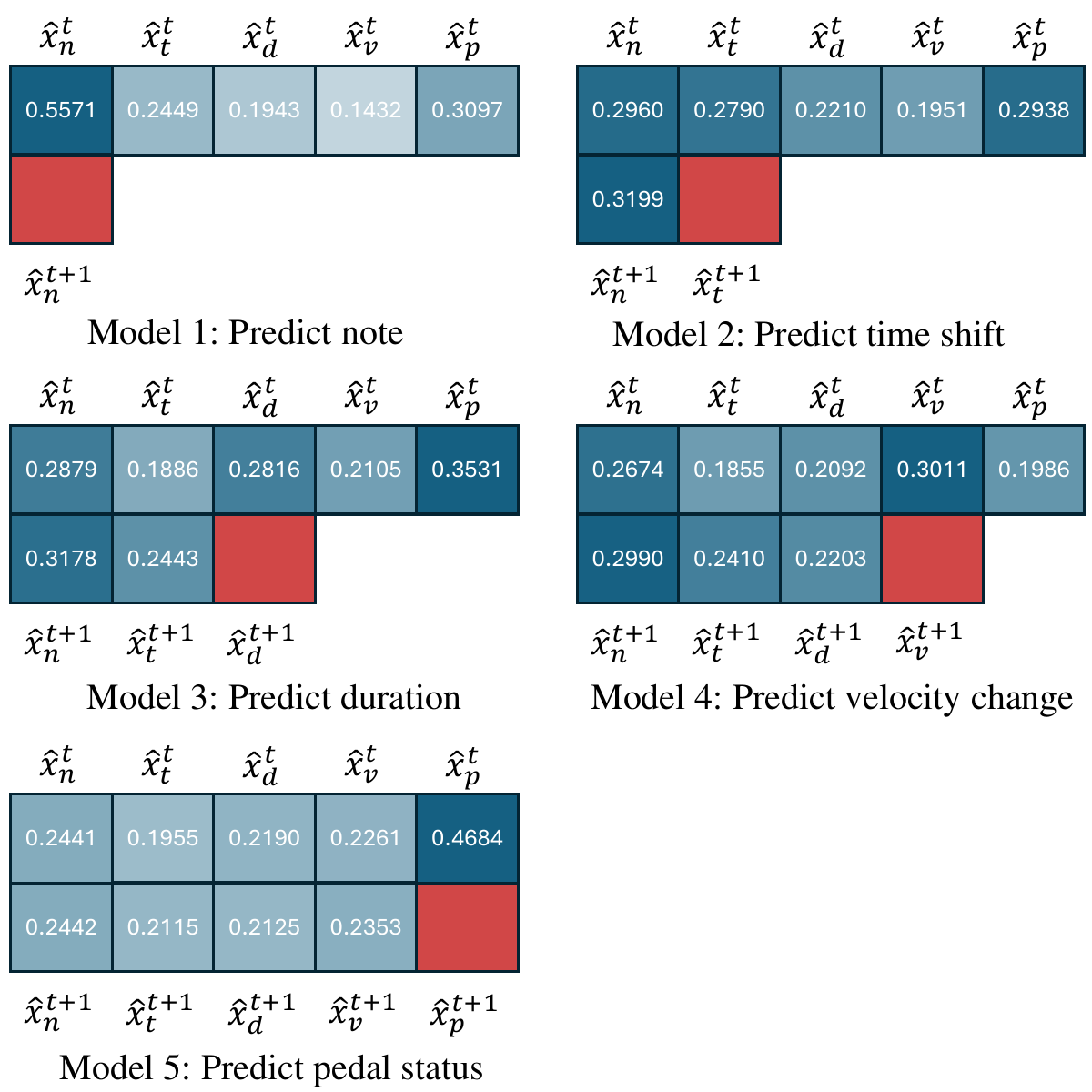}
\caption{Attention Score for Five Sequential Models. The weights are computed from the neural network parameters related to each input field of the models.} \label{fig: att}
\end{figure}

The weight distribution of each model w.r.t. its input fields can be summarized as attention scores in Figure \ref{fig: att}. We can make several instructive observations from it. First, the predictive field presents a trend of self-referencing. The weight block right above the red predictive block tends to be darker -- weighs heavier in predicting the output. In other words, when predicting the next note pitch, among all the input arguments, the pitch argument of the current note plays a heavier role. The self-referencing rule applies to other predictive fields as well. This observation well corresponds to our former assumption of musical element autonomy, namely the musical elements such as pitch and rhythm can be developed on their own at some level, with more attention on self-propagation which is less influenced by other elements.

A good analogy here is to view the five LSTMs as five independent channels propagating in time. Then the picture of sequential models is not only horizontal -- 1-dimensional in time, but also vertically layered with multiple channels advancing together (Figure \ref{fig: LSTM5}). When investigating this 2-dimensional process, we see that each channel is operating on its own with autoregressive self-referencing but at the same time, it also communicates with other channels by assimilating their outputs, both in the past and present. The communications with other channels are reflected by the attentional weights in fields other than the red predictive field in Figure \ref{fig: att}. Here we make a rough analogy of our model to the communicative channels and the attentional weights are an analogical measure of inter-channel communications. For a systematic investigation of channel communication using transfer entropy, please refer to this work \cite{dubnov2023switching}.

Another observation is on the weights of the current and previous arguments. The input of our models is conditioned on the sampled arguments of the current note from previous submodels. From the attention score, we can tell that for all five LSTMs, the weights of the current note are heavier than that of the input note. To be precise, the weights in the current note arguments are heavier than those in the same field in the input note (the second row weighs more than the first row). This attention distribution fits our intuition: to predict the duration of the current note, the pitch of the current note is more important than that of the previous note. This result further reveals the importance of modeling output interdependencies. If the output arguments are sampled independently, the second row of the attention score needs to be removed thus compromising the model performance with a loss of significant information.

A few specific observations can also be made from the attention score. We notice that the pitch argument $\hat{x}_n$ weighs heavier than other arguments in general, which shows that the note value plays an important role in predicting all musical fields. Another observation is that when predicting time-related musical features such as time shift and duration, the sustain pedal weighs heavily, which corresponds to the musical truth that the pedal sustains the effects of previously pressed notes over time.

\section{Selection with Entropy Sequence}\label{sec4}

At inference time, the trained LSTMs generate note sequences iteratively by feeding their sampled output as input to the models. These generated sequences are subject to \textit{exposure bias} -- when the model samples a value that's never observed in the data, all subsequent decisions of the model will be based on this alien value, which might cause error propagation and huge deviation of the generation from the ground truth.

As we are performing random sampling for each output, there is always a chance to sample alien values thus deteriorating the sequence. To avoid these deficient samples, we need to set criteria to select eligible sequences from all sampled generations.

\subsection{Entropy Sequence}

The intuition is that when the exposure bias happens, namely an unseen note is accidentally sampled and the subsequent generative process is disturbed, the system is expected to be confused thus behaving chaotically and producing more errors. The level of chaos, instability, or uncertainty is described by entropy. In other words, when the generative process is disturbed, its subsequent outputs should become more chaotic since the original order is broken and there is no established path to follow. Therefore, the output entropy will be higher than usual thus indicating a flawed sequence.

Using output entropy as a criterion to screen the sampled sequence is a subjective method, which is largely dependent on the training of the model rather than the ground truth. Let's say we have an ideal predictor that fits the data perfectly and doesn't generalize to any exterior samples, then we can select the valid sequences based on the output entropy since the exposure bias potentially causes system malfunction. However, for a neural network model with generalization ability, it's possible to generate a sequence out of the ground truth with high certainty (low entropy), while some ground truth samples may result in high entropy if they're not fitted well in training.

In other words, selection by output entropy is a model-based method but not a data-based one. With the trained model, we select the most stable and predictable sequence the model "thinks" with low output entropy. If the model is trained sufficiently well, this selection criterion generally stands. However, this relationship is not absolute. Lower entropy usually gives better sequences, but it doesn't mean the lowest entropy gives the best sequence. Many prevalent music sequences are variations of each other which provide substitutes for some outputs thus increasing the entropy. The sequence with the lowest entropy might inversely be some marginalized path with only one eligible note progression.

In our model, we perform the post-generation selection with low output entropy regulated with ground truth statistics. The regulation takes effect in two folds: on the one hand, it perturbs the order thus the system doesn't always choose the sequence with the lowest entropy, which might not be optimal; on the other hand, it compensates the training distortion of the data. Our neural network is not an ideal predictor and the entropy for ground truth may not be the lowest due to generalization and insufficient fitting. Regulating the generated sequences with data statistics selects generations with similar behaviors as the ground truth, which in turn improves the model performance.

Our evaluation of the overall sequence quality is similar to the \textit{reward} in reinforcement learning \cite{yu2017seqgan}. The difference is that we didn't train the model with reward but used it as a criterion to select from generations after training. Our proposed metric for sequence behavior can also be designed as a reward to train the model with reinforcement learning, which will be left as future work.

\begin{figure}[h]
\centering
\includegraphics[width=1\columnwidth]{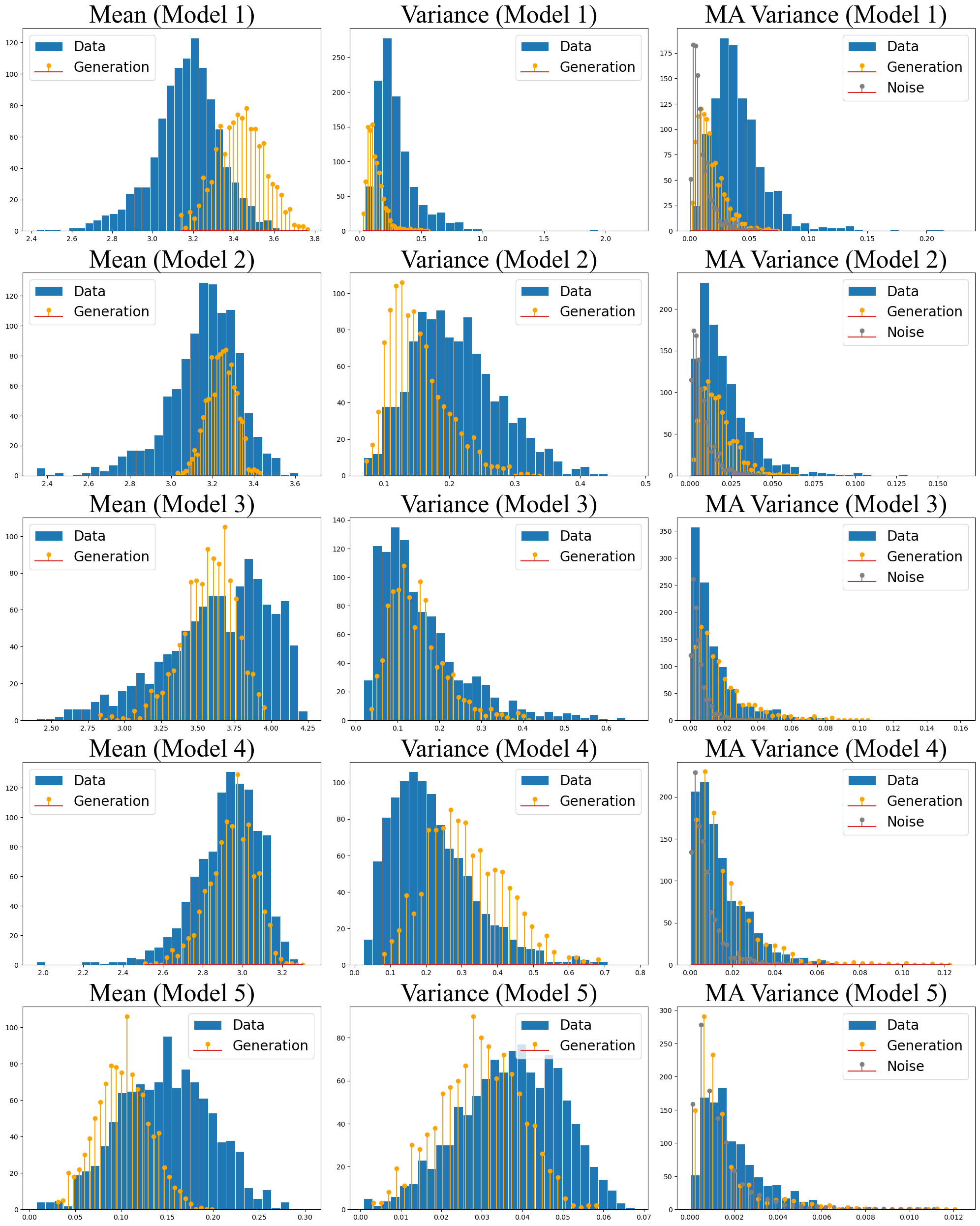}
\caption{Statistics of Entropy Sequence. This figure presents the mean, variance, and moving average variance distributions of the data and generation entropy sequences from the five sequential models in Figure \ref{fig: LSTM5}.}  \label{fig: stat}
\end{figure}

Notice that the output of a sequential model is a sequence, thus the output entropy is also a sequence. To select one eligible sample, we first sort all generations based on the average of their entropy sequence ascendingly, then regulate the order with ground truth statistics slightly and output the first sample in the regulated order. The statistics we compute for the entropy sequences include mean, variance, and moving average variance. The result is shown in Figure \ref{fig: stat}.

\subsection{Theoretical Corner}

To study the behavior of a sequence, we need to find suitable descriptors or metrics to quantify its qualities. As for the entropy sequence, the average or mean is a good indicator as discussed above, where low average entropy indicates a predictable and stable generation from the model, while high average entropy usually gives a chaotic and defective sample. Variance describes the sequence's deviation from the mean, which is a measurement of overall fluctuations. Mean and variance are classic statistical descriptors but they don't encode the trend of the sequence. We can generate noise signals with given mean and variance but the sequence is randomly ordered. Here we introduce the moving average variance as another metric to describe the progression of the sequence. The moving average smooths the small fluctuations of the sequence and shows its overall trend. The variance of this smoothed sequence better reflects the trend of fluctuations as the sequence unfolds in time.

The motivation for introducing variance and moving average variance in this problem is counter-intuitive. Generally speaking, if we aim to generate a sequence with the greatest stability and certainty, we should search with the least sequence mean and variance. However, in terms of music or arts in general, the aesthetic objective is not safety, certainty, or correctness as in engineering problems. Mistakes guide evolution and motivate change and creativity sometimes \cite{cascone2017aesthetics}. As for aesthetic measures, information theory and Kolmogorov complexity are widely used as metrics to gauge the trade-off between order and complexity in artworks \cite{rigau2008informational}. In music information dynamics, mutual information is proposed as a measure for musical expectations that encodes musical tendencies from the past to the present or future \cite{dubnov2021deep}. 

The motivation for using mutual information to measure musical pleasure comes from a trade-off between boredom and complexity. If there's little information encoded in the progression of a temporal musical event, the listener will soon find it boring or mundane, thus losing interest in listening. If too much new information is crammed into the process, the listener will find it difficult to comprehend thus also giving up listening. Finding the cognitive sweet point to engage the audience thus becomes the objective of musical aesthetics, which can be quantified by informatic mutual information \cite{abdallah2009information}.

The mutual information is defined as
\begin{equation*}
    I(X,Y) = H(Y) - H(Y|X)
\end{equation*}
which describes the information gain of $Y$ from $X$. The goal of maximizing musical pleasure or interestingness is to maximize the mutual information. However, to define $X$ and $Y$ computationally is difficult as this subjective measure involves the listener's cognition in determining the complexity. 

Intuitively, the mutual-informatic measure expresses an originally complicated situation of $Y$ (large $H(Y)$) being simplified with the presence of another variable $X$ (small $H(Y|X)$), which focuses on the amount of complexity reduction and the effect of $X$. In a temporal musical event, the composer can increase the systematic complexity by keeping the possibility open and reduce the complexity along the musical process by narrowing down the options and guiding listeners' expectations in a specific direction. In another way, when the musical tendency is too obvious (small $H(Y)$) and there's little room left to reduce entropy, the composer uses devices such as delay and suspense to build up tension (increase complexity) temporarily, thus realizing the maximal musical satisfaction at release by maximizing the effect of the final resolution in reducing musical uncertainty \cite{meyer2008emotion, meyer1957meaning}.

To capture this effect of maximal uncertainty reduction in our model, we design an analogical measure of mutual information, which is operable in predictive neural network models and easy to compute from the entropy sequence. In our problem, as all predictions of autoregressive neural networks are conditioned on the past, a division of $X, Y$ as past and present seems out of context since $H(Y)$ is difficult to compute. Alternatively, we treat $X$ as the larger context of the entire sequence. Therefore, the unconditioned complexity is represented by the in-the-moment output entropy, which is a quantity computed before the entire blueprint of the entropy sequence is laid out; the conditioned entropy is in turn a quantity that describes the overall behavior of the sequence, evaluating the sequence from a post-generated perspective.

Variance is a viable measure for the in-the-moment behavior of the entropy sequence, quantifying the fluctuations of the musical tendency. The sequence mean can serve as an overall indicator of the sequence quality after it finishes. Based on the informational aesthetic measure, here we should minimize the mean while maximizing the variance of the entropy sequence to realize the maximal musical satisfaction.

Interstingly, from Figure \ref{fig: stat} we can tell that when the generative mean is larger than the ground truth (model 1 \& 2), the variance tends to be smaller. It's far from drawing any conclusions as there are alternative explanations and also other factors at play. However, one thing we can tell is that the sequence variance is not the smaller the better -- the ground truth entropy sequences also fluctuate at some level in musical propagations. Besides, the moving average variance also tends to be large (compared to the noises), which shows certain regularities in the progression of the entropy sequence. In other words, the entropy sequences do not fluctuate randomly as white noise. Instead, there is a mindful mechanism to arrange these fluctuations, either being reduced linearly along the progression or temporarily increased and then reduced by suspense and release, as cross-referenced with the compositional techniques.

\bibliography{icmc2025_paper_template}

\end{document}